\documentstyle[pra,aps,preprint]{revtex}
\def\geqap{\,\raise 2pt \hbox{$>\kern-11pt \lower 5pt \hbox{$\sim$}$}\,}
\def\leqap{\,\raise 2pt \hbox{$<\kern-10pt \lower 5pt \hbox{$\sim$}$}\,}
\begin{document}
\draft
\title{Phase Diagram in R$_{1-x}$A$_x$MnO$_3$}
\author{Ryo Maezono, Sumio Ishihara$^{*}$, and  Naoto Nagaosa}
\address{Department of Applied Physics, University of Tokyo,
Bunkyo-ku, Tokyo 113, Japan}
\date{\today}
\maketitle
\begin{abstract}
We study the phase diagram of R$_{1-x}$A$_x$MnO$_3$ 
(R=La, Pr, Nd, Sm ; A=Ca, Sr, Ba)
by taking into account the degeneracy of $e_g$ orbitals and the
anisotropy of 
the transfer integral. 
The electron-electron interaction is treated in 
the mean field approximation with the optimization of the spin and
orbital 
structures. 
The global phase diagram is understood in terms of the 
two interactions, i.e., the super exchange interaction for small $x$ and
the double exchange interaction for larger $x$ modified by the orbital 
degeneracy.
The dimensionality of the electronic energy band resulting from  
the orbital structure is essential to determine the phase diagram. 
The effects of the Jahn-Teller distortion are also studied.
\end{abstract}


\narrowtext
The understanding of the rich phase diagram in 
R$_{1-x}$A$_x$MnO$_3$ (R=La, Pr, Nd, Sm ; A=Ca, Sr, Ba)
is indispensable for the discussion of its 
physical properties including the 
colossal magnetoresistance. 
These materials have been considered to be the model system of the 
double exchange mechanism \cite{zene,ande,dege,furu},
 i.e., the $t_{2g}$ spins are aligned
parallel in order to minimize the kinetic energy of $e_g$ electrons, 
which are strongly Hund coupled to $t_{2g}$ spins.
However this simple picture of the ferromagnetic phase has been
questioned
recently by several authors
\cite{mili,rode,hama,mizo,inou,ish1,kuko,ish2},
who stress the importance of the 
other interactions in addition to the double exchange one. 
The parent compound
LaMnO$_3$ is an insulator with the A-type antiferromagnetic (AF)
ordering and 
the Jahn-Teller (JT) distortion \cite{mats,woll}, while 
it should be metallic when only the Hund coupling is considered.
Roughly speaking there are two streams of thinking on this issue. 
One is to regard the JT distortion to be of the primary 
importance 
\cite{mili,rode}, 
which removes the degeneracy of the orbitals.
The other is to stress the strong 
correlation effects of the $e_g$ electrons 
\cite{inou,ish1,kuko,ish2}. 
In this picture the parent material is regarded as 
a Mott insulator, and the effective Hamiltonian is derived to 
study the spin and orbital structures \cite{ish1,kuko}. 
In the local density approximation (LDA) 
and LDA+$U$ band calculations for $x=0$ \cite{hama}, 
where the effect  of the electron correlation is included in 
a kind of mean field approximation,
it is concluded that the JT distortion of
the $(3x^2-r^2)/(3y^2-r^2)$-type is important for the 
A-type spin structure observed experimentally.
On the other hand, a recent exact diagonalization study of the 
effective Hamiltonian \cite{kosh} suggests that the correlation of the 
$(3x^2-r^2)/(3y^2-r^2)$-type or  $(z^2-x^2)/(y^2-z^2)$-type
orbital structure becomes remarkable in the A-type 
AF phase. 
%
%
As for the doped case $(x \ne 0)$, the system becomes ferromagnetic
metal for $x>0.175$ \cite{toku,schi}. 
The simple double exchange mechanism \cite{dege} 
is considerably modified as shown below when the 
anisotropy of the transfer integrals between the $e_g$ orbitals is taken 
into account. 
Especially it has been suggested that the orbital degrees
of freedom might remain disordered down to low temperatures
\cite{ish2}
to explain the anomalous physical properties.  
Hence the origin of the ferromagnetism 
should be reexamined taking into account the orbital degeneracy.
Near $x=0.5$ the charge ordering accompanied with the 
spin and orbital orderings has been observed.
With $x$ increased further  $(x \cong 0.6)$, the A-type
AF structure again appears, which shows quasi-two dimensional 
metallic behavior \cite{aki,kuwa}.
\par
In this paper we present an extensive study on the 
phase diagram of R$_{1-x}$A$_x$MnO$_3$ 
(R=La, Pr, Nd, Sm ; A=Ca, Sr, Ba) in the mean field approximation,
which treats both the super and double exchange 
interactions in a unified fashion at zero temperature.
Because the Coulomb interaction is the largest interaction and 
also the JT distortion disappears for $x>0.15$ \cite{kawa},
we first study the model 
with only electron-electron interactions.
The spin and orbital structures are optimized,
and the global phase diagram is given in the plane of $x$ ( the 
concentration of the holes )
and $J_s$ ( the super exchange interaction between the $t_{2g}$ spins ).
The effect of the JT distortion, which turns to be important for
$x=0$, is also studied.

We set up the three-dimensional cubic lattice consisting  of the 
manganese ions. 
Two kinds of the $e_g$ orbital( $\gamma, \gamma'$) 
are introduced on each site, 
and the $t_{2g}$ electrons are treated as a localized spin with
$S=3/2$.
The Hamiltonian without the JT coupling is given by \cite{ish1}, 
\begin{eqnarray}
H
&=&\sum_{<ij> , \sigma , \gamma , \gamma'}  
\Bigl( t^{\gamma \gamma'}_{ij} 
d_{i \gamma \sigma}^\dagger  
d_{j \gamma' \sigma}^{\phantom{\dagger}}  
+h.c. \Bigr) \nonumber \\
&+&
U\sum_{i \gamma} 
n_{i \gamma \uparrow} n_{i \gamma \downarrow} 
+ U'\sum_{i}  n_{i a} n_{i b} 
+ I \sum_{i,  \sigma , \sigma'} 
d_{i a \sigma}^\dagger  d_{i b \sigma'}^\dagger
d_{i a \sigma'}^{\phantom{\dagger}}   
d_{i b \sigma }^{\phantom{\dagger}}    \nonumber \\
&+& J_H \sum_{i} 
{\vec S_i} \cdot {\vec S^{t_{2g}}_i} 
+ J_s \sum_{<ij>} 
{\vec S^{t_{2g}}_i} \cdot {\vec S^{t_{2g}}_j} 
\ . 
\label{eq:ham}
\end{eqnarray}
$d_{i \gamma \sigma}^\dagger $ is the operator which 
creates an electron with spin $\sigma(=\uparrow, \downarrow)$ 
in orbital $\gamma(=a,b)$ at site $i$, and 
${\vec S_i}$ is the spin operator for the $e_g$ electron 
defined by 
${\vec S_i}=\frac{1}{2}\sum_{\sigma \sigma' \gamma} 
d^\dagger_{i \sigma \gamma} \vec \sigma_{\sigma \sigma'}
d_{i \sigma' \gamma} $.  
The electron transfer integral $t^{\gamma \gamma'}_{ij}$,
which is estimated by considering the oxygen $2p$ orbitals 
between the nearest Mn-Mn pair, 
is represented as  $c^{\gamma \gamma'}_{ij} t_0$, where 
$c^{\gamma \gamma'}_{ij}$ is the numerical factor depending 
on the orbitals and $t_0$ is estimated to be $0.72 eV$  
which we choose the unit of energy below ($t_0=1$) \cite{ish1}.
The second line shows the electron-electron interaction terms where
$U$, $U'$ and $I$
is the intra-, inter-orbital Coulomb interactions, and inter-orbital 
exchange interaction, respectively. 
This interaction can be rewritten as 
$
-\alpha \sum_i \Bigl(\vec S_i (\tau)
                      +\frac{J_H}{2 \alpha} \vec S_i^{t_{2g}}(\tau)
\Bigr)^2
  -\beta \sum_i \vec T_i (\tau)^2                 
$ \cite{ish2}.
Here the spin operator $\vec S_i$ and 
the iso-spin operator 
${\vec T_i}=\frac{1}{2}\sum_{\gamma \gamma' \sigma} 
d^\dagger_{i \sigma \gamma} \vec \sigma_{\gamma \gamma'}
d_{i \sigma \gamma'} $ for the orbital degrees of freedom
are introduced, and the two positive coefficients $\alpha$ and $\beta$, 
which are defined by 
$\alpha=2U/3+U'/3-I/6$ and $\beta=U'-I/2$ , represent the 
interaction to induce the spin and iso-spin moments, respectively. 
The last line is the sum of the Hund coupling 
and the AF interaction between the nearest neighboring 
$t_{2g}$ spins. 
Here we adopt the mean field approximation
by introducing the order parameters 
$ \langle \vec S_i \rangle $,
$ \langle \vec S^{t_{2g}}_i \rangle $,
and 
$ \langle \vec T_i \rangle$.
These order parameters are determined to optimize the mean field energy
at zero temperature.
For both spin and orbital, the four types of the ordering are
considered, 
that is, the ferromagnetic (F-type) ordering, where the order parameters 
are uniform, and the three AF-like  orderings, i.e., 
the layer-type (A-type), the rod-type (C-type) and the 
NaCl-type (G-type) AF orderings. 
Hereafter, types of the orderings are termed as, for example,
(spin:C), and so on. 
\par
In Fig. 1, the spin and orbital phase diagram 
is shown for the set of parameters $\alpha=70>>\beta=2.5$.
In this rather extreme case
the regions dominated by the super exchange and double exchange 
interactions are separated, and it is easy to obtain the physical
picture. 
$\alpha/\beta>>1$ corresponds to the situation where  
$1/(U'-I) >> 1/U, 1/(U'+I)$. 
In Fig. 1, 
the spin structure changes as F $\to $ A $\to $ C $\to$ G, as 
$J_s$ increases, which is consistent with the exact diagonalization
study 
\cite{kosh}.
We begin with the discussion of  the parent material ($x=0$) where only
the 
super exchange interaction is effective. 
For spin:A, which is observed in $\rm LaMnO_3$,  
the most stable orbital structure is the orbital:G
 $([3z^2-r^2] + [x^2-y^2])/([3z^2-r^2] - [x^2-y^2])$
as shown in Fig. 1.
When the ratio $\alpha/\beta$ is changed, 
the orbital changes continuously from 
orbital:G $([3z^2-r^2]+[x^2-y^2])/([3z^2-r^2]-[x^2-y^2])$ 
for $\alpha/\beta>>1$ to 
orbital:G $(y^2-z^2)/(z^2-x^2)$ 
for $\alpha/\beta \sim 1$, 
and  to  $(3z^2-r^2)$ 
for $\alpha/\beta=0$.
For the actual compound,  
we expect $\alpha \geqap \beta$ and  
orbital:G$(y^2-z^2)/(z^2-x^2)$ is the  most stable.
The experimentally observed 
orbital:G$(3x^2-r^2)/(3y^2-r^2)$ 
is never the most stable solution, which can be understood as follows.
There are three possibilities for the intermediate states of the 
super exchange processes, i.e., the occupancy of the two orbitals 
(a) with the parallel spins (the energy $U'-I$) or (b)
antiparallel spins ($U'+I$), and 
(c) the double occupancies of the same orbital ($U$) 
\cite{ish1,kuko,cyro,ina}.
Then the relative importance of the states (a) compared with (b) and (c)
is changed when $\alpha/\beta$ is changed.
Let us compare  the energy gains due to the 
super exchange processes in orbital:G $(y^2-z^2)/(z^2-x^2)$ and 
orbital:G $(3x^2-r^2)/(3y^2-r^2)$.
For the processes using states (a) and (b), the magnitudes of the 
transfer integrals and hence the energy gain are the same,
while for the process using (c), 
the energy gain is always larger 
for $(y^2-z^2)/(z^2-x^2)$ compared with $(3x^2-r^2)/(3y^2-r^2)$.
Then there is no chance for $(3x^2-r^2)/(3y^2-r^2)$ to be the 
most stable structure for any value of $\alpha/\beta$.
Hence the JT coupling is important in addition to the 
electron-electron interactions at $x=0$. 
We introduce the 
JT distortion observed experimentally and its 
coupling to the $e_g$ electrons. 
We consider the two shorter Mn-O bonds and the one longer bond 
in the $\rm MnO_6$ octahedron, and its bond length is represented as 
$d_{long}=d_0(1+0.056)$ and $d_{short}=d_0(1-0.028)$, respectively, 
as we follow the structural data \cite{mats}. 
The change of the transfer integrals is estimated in terms of the 
dependence of the $3d-2p$ hopping $t_{pd}$ on the distance $d$ as
$t_{pd}  \propto d^{-7/2}$ \cite{harr}. 
We vary the splitting of the energies ($g$)
between the two orbitals as the parameter and found

\noindent
[1] The wave functions are saturated to become 
$(3x^2-r^2)/(3y^2-r^2)$ when $g$ is about the half of the transfer
energy $t_0$.
This value is much smaller than what is expected 
in the absence of the electron-electron interactions.
The magnitude of the isospin moment $|\vec T_i|$ is already induced
almost fully in terms of the strong electron-electron interactions,
and the role of the JT coupling is to fix the direction of  
$\vec T_i$. 
\noindent
[2] The spin:A is stabilized relative to spin:F by JT distortion. 
The phase boundary $J_s(FA)$ between A and F is shifted from  
$J_s(FA) = 0.014$ for $g=0$ to $J_s=0.007$ for $g=1.0$.
This tendency is in agreement with the other calculations
\cite{hama,mizo}, but the physics is different.  
In the band calculation, the ground state without the JT distortion 
is the ferromagnetic metal and the enhanced AF exchange between layers 
is due to the reduced double exchange interaction by 
JT distortion \cite{hama}. 
In our calculation, on the 
other hand, only the super exchange interactions are relevant because
the large gap has been already opened up due to the strong
electron-electron interactions. 
The stabilization of the A-AF is clearly understood in terms of the 
change of the super exchange interaction by the 
orbital rearrangement. 
\par
Now let us turn to the doped case ($x \ne 0$).
The orbital structure in spin:F 
is quite sensitive to the carrier concentration,  
that is, it changes continuously as $x$ increases from
orbital:G$(x^2-y^2)/(3z^2-r^2)$ near $x=0$ to $(x^2-y^2)$ for $x \cong 0.3$,
and to orbital:A$([3z^2-r^2]+[x^2-y^2])/([3z^2-r^2]-[x^2-y^2])$ 
for $0.3<x<0.8$ and finally $(3z^2-r^2)$ for $x=0.8$, as shown in Fig.1. 
On the other hand the orbital in spin:A and spin:C almost remains 
$(x^2-y^2)$ and $(3z^2-r^2)$, respectively, expect for $x=0$,  
in contrast to spin:F case. 
The phase boundary $J_s(FA)$ increases linearly near 
$x=0$, and turns to decrease to have a minimum $J_s(FA) \cong 0$
at around $x=0.3$, where both spin:F and A have the $(x^2-y^2)$ orbital. 
The linear increase is due to the difference in the location of the 
band edges for spin:F and spin:A structures. This feature
remains true even when the canting in the spin:A is taken into account because
it gives the energy gain only of the order of $\sim x^2$ \cite{dege}. 
The minimum of $J_s(FA)$ around $x \cong 0.3$, 
separates rather clearly the two regions dominated by the super 
exchange ($x \leqap 0.3$)
and the double exchange interactions ($x \geqap 0.3$).  
In the doped case the shape of the density of states and 
the Fermi energy is crucial to determine the double exchange 
energy, which depends on both spin and orbital structures.
Especially The dimensionality and the van-Hove singularities 
of the density of states depends strongly on the orbital 
structure. 
Therefore, in that sense, the double exchange mechanism is 
considerably modified from the conventional one, 
when the anisotropy of the transfer integrals 
and the electron-electron interaction are taken into account. 
In the region of $x < 0.3$,  
the orbital:G$(x^2-y^2)/(3z^2-r^2)$- and orbital:F$(x^2-y^2)$-type
structures realized in the spin:F and spin:A phases respectively, and 
the band becomes two-dimensional. 
Here, the density of states has a logarithmic singularity 
at $\varepsilon=0$ and decreases monotonously 
as $|\varepsilon|$ with the steps at 
$\varepsilon= \pm 3t_0$.
In the low carrier concentration 
region, the above type of the density of states 
is more favorable than the three dimensional one due to the step 
at the band edge. 
In this case there is no difference in the kinetic energy 
of the carriers between spin:F and spin:A, and 
$J_s$ favors spin:A. 
Then the F region in Fig. 1 for $x<0.3$ is 
stabilized by the super exchange interaction. 
In the region of $x> 0.3$, 
the orbital structure is rearranged as 
orbital:A$([3z^2-r^2]+[x^2-y^2])/([3z^2-r^2]-[x^2-y^2])$-type, 
where the band structure is essentially three dimensional, 
but the density of states has two peaks at $\varepsilon \cong \pm 2t_0$
and resembles that of the one-dimensional band.  
Eventually the $(3z^2-r^2)$ orbital appears at $x=0.8$ and  
gives the one dimensional-like band along the 
$z$-axis where the density of states has the peak structures at
$\varepsilon =\pm 3t_0$.
Then the adjusting of the orbital structure with increasing $x$ 
occurs in order to 
minimize the kinetic energy, i.e., the 
center of mass for the occupied states. 
When one consider the occupied orbital, the energy
position of the band edge does not depend on the dimensionality
of the dispersion.
Then only the shape of the density of states matters, and 
one dimensional-like dispersion is advantageous in this viewpoint.
For spin:C, the orbital is 
$(3z^2-r^2)$ almost always except at very small $x$.
This can be easily understood because the spin structure allows the
electron motion only along the $z$-axis.
For spin:G, the electron motion is blocked in all 
directions and the double exchange energy gain is absent.
Then the electronic energy does not depend on the orbital structure
in the limit of strong electron-electron interaction.
\par
In Fig. 2, we present the calculated phase diagram in the case 
of $\alpha = 8.1$, $\beta=2.5$, which is more relevant to the 
actual manganese oxides. This set of parameters is complementary to 
that in Fig. 1 because both $\alpha/\beta$ and  $\alpha$ are
smaller.
In comparison with the results in Fig. 1, 
the spin:F region dominated by the super exchange interaction, 
is extended to the region with larger $x$. 
This results in the merging of the super exchange and double 
exchange regions.
First consider the spin:F state.
At $x=0.0$, the orbital structure in the spin:F phase is 
the same as that in Fig. 1. 
What is different from Fig. 1 is that as $x$ is increased 
the orbital  becomes 
orbital:C $([3z^2-r^2]+[x^2-y^2])$/$([3z^2-r^2]-[x^2-y^2])$-like
structure
instead of $(x^2-y^2)$ 
and to the 
orbital:A $([3z^2-r^2]+[x^2-y^2])$/$([3z^2-r^2]-[x^2-y^2])$ around
$x=0.6$.
The two dimensional $(x^2-y^2)$-like structure appears for $x \cong 0.9$ 
where the mixing of the two bands is important.

As for the spin:A, C,  G, the orbital remains 
basically the same as in Fig. 1.
Then it is concluded that the orbital structure is sensitive to the 
interactions only in spin:F. This is related to the 
the degeneracy of the orbital structures \cite{ish2}.
Actually all the obtained orbital structures in Figs. 1 and 2 belong
to the lowest degenerate states discussed in \cite{ish2}.
\par
We now discuss the comparison between the mean field phase diagram in
Fig. 2 
and the experiments. 
In the present mean field calculation, 
the ferromagnetic phase, i.e. the spin:F phase is growing up 
with increasing $x$ from the insulating state, and it becomes
most remarkable around $x=0.3$, as shown in Fig. 2. 
The global feature of the spin:F phase is 
quite consistent with the experimental 
results in $\rm La_{1-x}Sr_xMnO_3$, where the ferromagnetic phase
appears at about $x=0.08$ and it survives up to $x=0.5$. 
It is worth to note that, however, 
the origin of the ferromagnetic phase is far from the 
conventional double exchange mechanism, i.e.,
both the super exchange and the double exchange interactions 
modified by the  orbitals are relevant in the region $0.2<x<0.4$.
The orbital ordering in spin:F, if observed experimentally by neutron
and/or X-ray diffraction, will give important clues to the interactions
because it depends sensitively on the parameters as described above.
The another possibility is the orbital fluctuation
is so large that the orbital liquid state is realized \cite{ish2}. 
The RPA analysis of the mean field solutions is left for 
the future work. 
\par
The another implication to the experimental results is
about spin:A phase appearing around $x>0.5$. 
In $\rm Nd_{1-x}Sr_xMnO_3$,
the ferromagnetic metallic phase is realized up to about $x=0.48$ and 
the CE-type AF structure with the charge ordering tunes up 
\cite{aki,kuwa}. 
With further increasing of $x$, 
the metallic state with spin:A again appears at about x=0.53,  
and the large anisotropy in the electrical resistivity is observed 
in this phase. 
The similar metallic phase accompanied with spin:A is also reported in 
$\rm Pr_{1-x}Sr_xMnO_3 $ \cite{kuwa}. 
Although the charge ordered phase is not considered, i.e., 
the long range Coulomb interaction is neglected, in the present work, 
the global phase change, as 
spin:A insulator ($x \sim 0$) $\rightarrow$
spin:F metal ($ 0.1 \leqap  x  \leqap 0.5$) $\rightarrow$
spin:A metal ($ 0.55 \leqap x $), 
is well reproduced when we fix $J_s$ to be around $0.02eV$,
which is a reasonable value, as shown by the broken line in Fig. 2. 
It is predicted that the $(x^2-y^2)$-type orbital structure 
should be realized in this spin:A metallic phase, in contrast 
to the insulating phase with spin:A appearing at $x=0.0$.  
Furthermore, 
the AF interaction between layers is expected to be  
enhanced in comparison with that at $x=0.0$, 
because the ferromagnetic interactions 
originated from the both double exchange and super exchange interactions 
are prohibited in this direction.
By the same reason it is predicted that the spin canting does not occur 
in this spin:A metal as observed in \cite{kuwa}, 
in contrast to the spin:A in the small $x$ region
which has been discussed by de Gennes \cite{dege} 
\par
In summary we have studied the phase diagram of La$_{1-x}$Sr$_x$MnO$_3$
in the plane of $x$ ( hole concentration ) and 
$J_s$ ( AF exchange interaction between the $t_{2g}$
spins) in the mean field approximation. 
The global features can be understood in terms of the interplay between
the super exchange and the double exchange interactions which 
are considerably modified with taking the orbital degrees 
of freedom into account. 
The dimensionality of the energy band 
attributed to the orbital structure plays 
essential roles to determine the phase diagram. 
The orbital structure  
is sensitive to changes of the carrier concentration and the 
interactions only in the ferromagnetic state, which 
suggests the importance of the ordering/disordering of the orbital on 
the origin of the ferromagnetism in the perovskite manganites. 
\par
The authors would like to thank S. Maekawa,  
Y. Tokura, K.Terakura, and I.Solovyev for their valuable discussions. 
This work was supported by the Center of Excellence Project
from the Ministry of Education, Science and Culture of Japan, 
and the New Energy and Industrial Technology Development 
Organization (NEDO).  
\par
\noindent
$^{*}$  Present address: 
Institute for Materials Research, Tohoku University, 
Sendai, 980-77, JAPAN. 
\vfill 
\eject
\noindent
Figure captions
\par
\noindent

Figure 1. The mean field phase diagram 
in the plane of the carrier concentration $(x)$ and the 
antiferromagnetic interaction $J_s$
between the $t_{2g}$ spins. 
The strength of the interactions are set as $\alpha=70 >> \beta=2.5 $.
The schematic orbital structure 
in the each phase is also shown. 
\par
\noindent
Figure 2. The calculated mean field phase diagram 
with $\alpha = 8.1$ and $\beta=2.5$ case. 
\par \noindent
\end{document}